\documentclass[lettersize,journal]{IEEEtran}
\usepackage{amsmath,amsfonts}
\usepackage{algorithmic}
\usepackage{algorithm}
\usepackage{array}
\usepackage[caption=false,font=normalsize,labelfont=sf,textfont=sf]{subfig}
\usepackage{textcomp}
\usepackage{stfloats}
\usepackage{url}
\usepackage{verbatim}
\usepackage{graphicx}
\usepackage{cite}
\usepackage{xcolor}
\makeatletter

\newcommand{\Rmnum}[1]{\expandafter\@slowromancap\romannumeral #1@}
\makeatother

\begin{document}

\title{Sim-T: Simplify the Transformer Network by Multiplexing Technique for Speech Recognition}

\author{Guangyong Wei, Zhikui Duan, Shiren Li, Guangguang Yang, Xinmei Yu, Junhua Li
\thanks{Guangyong Wei, Zhikui Duan, Xinmei Yu, Guangguang Yang and Junhua Li are from Foshan University, Foshan city, Guangdong, P.R.China. (e-mail: gy.wei@foxmail.com; duanzhikui@outlook.com; labxmyu@fosu.edu.cn; sunshineuop@163.com; jh.io@foxmail.com)}
\thanks{Shiren Li is from Sun Yat-Sen University, Guangzhou, P.R.China. (e-mail: lishr6@mail3.sysu.edu.cn)}
\thanks{Corresponding author: Zhikui Duan}}

\markboth{Journal of \LaTeX\ Class Files,~Vol.~14, No.~8, August~2023}%
{Shell \MakeLowercase{\textit{et al.}}: A Sample Article Using IEEEtran.cls for IEEE Journals}


\maketitle

\begin{abstract}
In recent years, a great deal of attention has been paid to the Transformer network for speech recognition tasks due to its excellent model performance. However, the Transformer network always involves heavy computation and large number of parameters, causing serious deployment problems in devices with limited computation sources or storage memory. In this paper, a new lightweight model called Sim-T has been  proposed to expand the generality of the Transformer model. Under the help of the newly developed multiplexing technique, the Sim-T can efficiently compress the model with negligible sacrifice on its performance. To be more precise, the proposed technique includes two parts, that are, module weight multiplexing and attention score multiplexing. Moreover, a novel decoder structure has been proposed to facilitate the attention score multiplexing. Extensive experiments have been conducted to validate the effectiveness of Sim-T. In Aishell-1 dataset, when the proposed Sim-T is 48\% parameter less than the baseline Transformer, 0.4\% CER improvement can be obtained. Alternatively,  69\% parameter reduction can be achieved if the Sim-T gives the same performance as the baseline Transformer. With regard to the HKUST and WSJ \textit{eval92} datasets, CER and WER will be improved by 0.3\% and 0.2\%, respectively, when  parameters in Sim-T  are 40\% less than the baseline Transformer.
\end{abstract}


\section{Introduction}
\IEEEPARstart{T}{ransformer}\cite{transformer2017} network has been widely applied in several application scenarios, such as natural language processing (NLP)  \cite{Bert, ALBert, Q8Bert}, vision tasks \cite{EAPT, liu2021swin, fan2021multiscale}, language model (LM) \cite{transformer_XL, transformer_LM} and automatic speech recognition (ASR) \cite{speech_transformer, LFEformer, gulati2020conformer, he2020realformer, lstransformer, aes}. In particular, Transformer and its variants provide better performance than other network models in the field of ASR with the basis of the combination of attention mechanism and feedforward network (FFN). However, the Transformer-based models tend to suffer from a large number of parameters and heavy computation. 

\begin{figure*}

  \centering
  \includegraphics[width=6.5in]{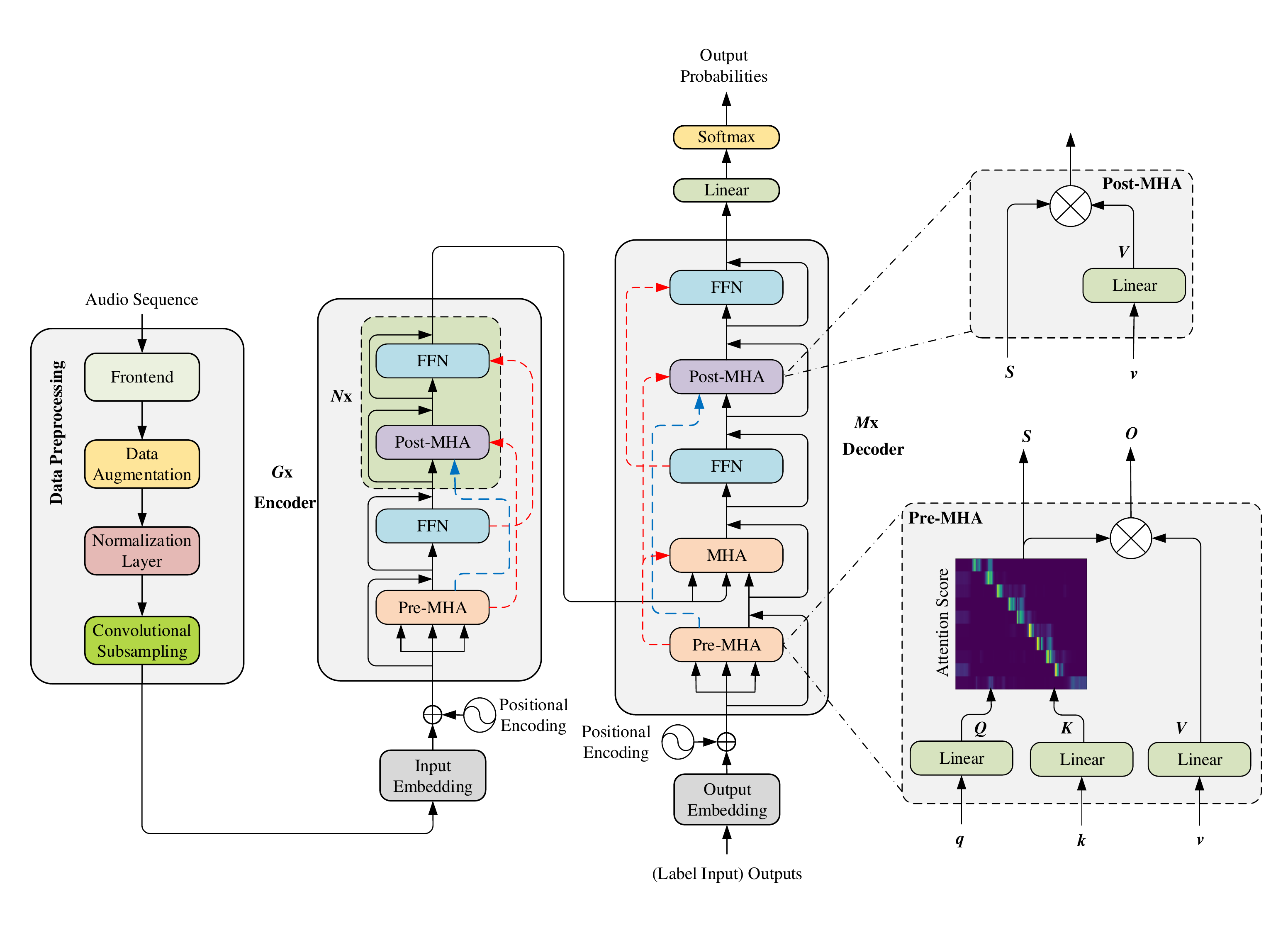}
  \caption{This is the overall structure of the proposed Sim-T network, including three parts: data preprocessing, encoder, and decoder. \textit{$G\times$} means that all encoding layers are divided into $G$ groups. The red dotted line indicates the module weight multiplexing technique. The blue dotted line represents attention score multiplexing.}
  
\label{fig_model}   
\end{figure*}

The huge amount of parameters requires sufficient storage and memory, making these high-performance models unsuitable for devices with limited computing resources. 
This, therefore, makes model compression become a desirable solution to the mentioned problems.
Weight sharing, originally applied in neural networks by \cite{lecun1989generalization} and \cite{nowlan2018simplifying}, has been demonstrated as an effective way to compress models. Recently, \cite{dehghani2018universal} applied the idea of weight sharing to the Transformer network for model compression in the field of machine translation and LM. Following that, the scheme of cross-layer parameter sharing was put forward in the ALBert network \cite{ALBert}, where the weights of the corresponding modules from all layers are shared. As mentioned in \cite{Redundancy}, the features tend to become increasingly different as the difference in their layer index increases. Accordingly, features from adjacent layers are similar. Under this circumstance, sharing weights of corresponding modules from all the network layers, such as ALBert network \cite{ALBert}, is not the optimal solution. Moreover, the effective reduction of parameters using weight sharing has not been fully explored in the field of ASR. In order to effectively address the module compression problem in ASR tasks, a new scheme called module weight multiplexing has been proposed. In this scheme, all network layers are first divided into several groups, and then weights from the corresponding modules of each group are multiplexed. As shown in the red dotted line from Fig. \ref{fig_model}, multiple FFN modules and various attention modules perform weight multiplexing within each group or layer. The proposed scheme can systematically explore how the number of divided groups actually imposes an effect on model performance and the number of model parameters.




According to the open literature, except for the proposed module weight multiplexing, attention score multiplexing has never been explored before though it is also useful for model compression in the Transformer-based network. It is worth mentioning that in the Transformer network, the attention scores used for representing the global interaction among different positions of the input tokens greatly suffer from the quadratic computational complexity of the inputs. This is tempted to take up a lot of computing resources. \cite{shim2021understanding} studied the redundancy of attention matrices based on the degree of diagonalization for which high similarity of the cumulative attention diagonals between adjacent layers can be noticed. 
From this perspective, attention score multiplexing is useful for model compression and computation overhead reduction in Transformer-based models. 
Specifically, for each group, the attention score is only generated in the first attention module and then shared across all the subsequent attention modules.
At the same time, the label sequences are used as the input of the first decoder layer, and their context information is continuously weakened as the model goes deeper. This issue can be well addressed by the proposed attention score multiplexing. However, the attention score multiplexing technique can not be directly inserted into the Transformer decoder layer. To solve this problem, a new decoder layer, which is suitable for the attention score multiplexing technique, has been developed. It is salient that the same performance can be achieved with fewer newly proposed decoder layers than the original Transformer decoder layers. In this sense, the model parameters can also be reduced.

In this study, a lightweight network called Sim-T has been proposed for which the multiplexing technique is used for simplifying the Speech-Transformer network \cite{speech_transformer}.
The proposed multiplexing technique consists of two parts: module weight multiplexing and attention score multiplexing. For the encoder, we divide the network layers into multiple groups and apply both module weight multiplexing and attention score multiplexing to each group. The module weights and attention scores are initially updated by the first layer of each group, then they are reused by other layers of the same group. For the decoder, this module weight multiplexing technique is applied to each network layer, and the attention scores of the labels are reused in each network layer under the help of the attention score multiplexing. The contributions of this paper can be summarized as follows:

\begin{itemize}
    \item A lightweight network called Sim-T has been proposed for ASR tasks by employing the newly developed module weight multiplexing and attention score multiplexing technique.
    \item A novel decoder structure has been proposed to enhance the label information and enable the proposed attention score multiplexing to be directly applied in the speech Transformer network.

      \item Promising results can be achieved from Sim-T. In Aishell-1 dataset, when the proposed Sim-T is 48\% parameter less than the baseline Transformer, 0.4\% CER improvement can be obtained. When the Sim-T has the same performance as the baseline Transformer, 69\% parameter reduction can be achieved. In HKUST dataset and WSJ \textit{eval92} dataset, when the proposed Sim-T is 40\% parameter less than the baseline Transformer, CER and WER will be improved by 0.3\% and 0.2\%, respectively.
\end{itemize}

\section{Related Work}
In the field of ASR, extensive attention has been paid to end-to-end networks \cite{tmm_speech} in recent years, such as CTC \cite{CTC} and Encoder-Decoder models. In particular, the Encoder-Decoder models have been widely studied and a lot of relevant works have been published, such as RNN \cite{RNN}, LSTM \cite{LSTM}, LAS \cite{LAS} and Transformer \cite{transformer2017}.



Transformer, whose excellent model performance mainly originates from the combination of the self-attention mechanism and feed-forward network, was originally proposed in the translation task and later favored in the field of ASR.
Speech-Transformer \cite{speech_transformer}, as a good example, is an application of the pure Transformer framework. With the inspiration of the speech-Transformer network, lots of subsequent research \cite{LFEformer, gulati2020conformer, he2020realformer, lstransformer} have been conducted.

Model compression is a popular research field. \cite{Redundancy} and \cite{kovaleva2019revealing} have shown that there is a large amount of redundant information in the deep learning-based model. The parameters that play a decisive role in the model performance account for a small proportion. 
In addition, model compression can remove redundant information and calculation cost with almost no sacrifice of model performance.
Hence, the lightweight model has attracted extensive attention from researchers. 
Cases in point are demonstrated as follows. \cite{Q8Bert} adopted a quantization-aware training method to compress the model size and speed up the inference. \cite{Tinybert} and \cite{DeiT} used knowledge distillation approach to reduce model parameters. \cite{Dynamicvit} thinned the input tokens of each layer of the Transformer network to reduce the redundant information. ALBert\cite{ALBert} employed parameter decomposition and cross-layer parameter sharing to fully compress the model size, which can prevent parameters from growing with the increasing depth of the network layers.


The idea of weight sharing was originally introduced into neural networks by \cite{lecun1989generalization} and \cite{nowlan2018simplifying} in which a simple and effective method for model compression has been proposed. At present, some excellent works based on weight sharing, such as \cite{ALBert, dehghani2018universal, Leve_transformer, Minivit}, have been explored to address the problem of heavy computation and large parameter amount that are suffered by the Transformer network. In particular, Universal Transformer \cite{dehghani2018universal} used weight sharing to  the network layer, so as to achieve the effect of the RNN-like recurrent network. 
In addition, the Universal Transformer is different from the original Transformer since the number of encoder layers and decoder layers in the original Transformer network is fixed.
ALBert \cite{ALBert} used cross-layer parameter sharing to prevent the model size from increasing with the deepening of the network layers. Both Universal Transformer and ALBert share all module parameters across layers by default. Our proposed Sim-T systematically explores the impact of weight sharing applied at different depths of network layers on both model parameters and model performance. 
Considering the high computational complexity of the attention score in the Transformer network, we further expand the scope of weight multiplexing to the attention score multiplexing for model compression.

\section{Method}
In this section, the overall structure of our proposed Sim-T model will be introduced, including data pre-processing, encoder, and decoder. 
In addition, the principles of module weight multiplexing and attention score multiplexing will be explained.


\subsection{Overall Structure of Sim-T}\label{sec2_1}
The end-to-end Transformer \cite{speech_transformer} model shows excellent performance in the field of ASR. Taking advantage of the multi-head attention (MHA) mechanism, Transformer-based model can capture long-range context interaction relationships and extract high-level representations through feedforward networks. The proposed Sim-T network is a simplified and lightweight Transformer-based network, so they have similar advantages. The overall structure is shown in Fig. \ref{fig_model}. 
Audio features obtained through data preprocessing are input into the encoder, allowing the high-level acoustic features to be well achieved.
The output of the encoder and the label features are used for decoding through the decoder.

{\bf Data Preprocessing:} In the Sim-T model, data preprocessing is first performed on the input audio sequences, as shown on the left side of Fig. \ref{fig_model}. The original audio sequences are transformed from the time domain to the frequency domain via the front-end module to obtain log-Mel features. Then, time warp, frequency masking and time masking are applied to the log-Mel spectrogram through the Data Enhancement module, which can force the model to learn more robust features and improve its generalization ability. The features are then normalized using mean-variance in the Normalization Layer. 
The input features are downsampled by the convolutional downsampling module so that the computation amount in the subsequent procedures can be significantly reduced.
The downsampled features are added to the positional information and their summation is the input of the encoder layer.

{\bf Encoder:} The encoder is used to extract high-level acoustic features from the audio features after data preprocessing. The encoder consists of $L$ layers.  In the proposed Sim-T model, in order to analyze how the multiplexing technology affects the model performance and parameter reduction at different numbers of weight-sharing layers, we divide $L$ encoder layers into $G$ groups $(1 \leq G \leq L)$ for investigation. 
In particular, the number of layers in each group is defined as $L/G$. The weights in corresponding modules of each group are multiplexing rather than multiplexing all layers of the network like the ALBert network. As shown in the middle part of Fig. \ref{fig_model}, each encoder group mainly includes three modules: Pre-MHA, Post-MHA and FFN. Each module includes residual connection \cite{He2016DeepRL} and layer normalization \cite{Ba2016LayerN}. The first encoder layer in each group constructs the new module weights and attention scores, and the subsequent $N$ encoder layers reuse them,  where $N=L/G-1$. It is worth mentioning that when $G=L$, it means the encoder in Sim-T is identical to that in the original Transformer encoder. That is to say, no weight multiplexing is applied. $G=1$ means that all encoder layers share the same weights.

Given that the encoder input features can be represented as $x\in \mathbb{R}^{T_{1}\times d_{model}}$. ($T_{1}$ is the length of the acoustic feature; $d_{model}$ is the dimension of the input feature.) 
$X_{i}\in(X_{1},.., X_{G})$ is used to represent the output of $i$-th encoder group. $X_{i}^{j}$ is used to represent the output of the $j$-th layer in the $i$-th group. $X_{i}^{j}$ can be expressed as follows:

\begin{equation}
X_{i}^{j} =
\begin{cases}
FFN(O_{i})& \text{$j = 1$},\\
FFN[Post\mbox{-}MHA(S_{i},X_{i}^{j-1})]& \text{$1 \textless j \leq (N+1)$},
\end{cases}
\end{equation}

\begin{equation}\label{eq_encoder_LEMHA_out}
    S_{i}, O_{i}=Pre\mbox{-}MHA(X_{i-1}^{N+1}, X_{i-1}^{N+1}, X_{i-1}^{N+1}),
\end{equation}
where $X_{0}=x$. $S_{i}\in \mathbb{R}^{T_{1}\times T_{1}}$ and $O_{i}\in \mathbb{R}^{T_{1}\times d_{model}}$  denote the attention score and output feature of the Pre-MHA module, respectively, in the $i$-th group.

The traditional Transformer network uses a MHA to capture the global context information of the input sequences. The attention scores produced by both vector dot product and $Softmax$ can effectively represent this mentioned global content. 
However, the computation of attention scores is quadratic to the length of the input features. This means it suffers from heavy computation overhead and a large number of parameters.
Moreover, the attention scores from adjacent layers are similar, demonstrating that there is high redundancy between them. Thus, it is not an optimal solution to update the attention scores in each layer.

Here we propose two attention approaches, namely Pre-MHA and Post-MHA, shown in Fig. \ref{fig_model}. Pre-MHA is similar to the traditional attention mechanism for which the query vector ($q$), key vector ($k$) and value vector ($v$) are used as input and then mapped to $Q$, $K$ and $V$, respectively, through three different linear layers, so that the results of attention scores ($S$) and output feature ($O$) can be finally obtained. 
In particular, Post-MHA will share the attention scores calculated by Pre-MHA and dot product operations are conducted with the current input and the shared attention scores to reduce the number of updates in attention scores calculation and overall calculation cost.
The expressions of Pre-MHA and Post-MHA are as follows:
\begin{equation}\label{Pre-MHA}
    S, O=Pre\mbox{-}MHA(q, k, v),
\end{equation}
\begin{equation}\label{eq_mutli_attn}
    Pre\mbox{-}MHA(q, k, v) = Concat(Head_1,...,Head_h)W^o,
\end{equation}
\begin{equation}\label{head}
     Head_n = ATTENTION(Q,K,V),
\end{equation}
\begin{equation}\label{attn}
    ATTENTION(Q,K,V) =SV,
\end{equation}
\begin{equation}\label{W}
    S= SOFTMAX(\frac{{Q}{K^{T}}}{\sqrt{d_k}}),
\end{equation}
\begin{equation}\label{QKV}
    Q=qW_{n}^{q},K=kW_{n}^{k},V=vW_{n}^{v},
\end{equation}
\begin{equation}\label{Post-MHA}
    Post\mbox{-}MHA(S, v)=SvW_{n}^{v},
\end{equation}
where $W^{o}\in \mathbb{R}^{hd_{q}\times d_{model}}$, $W_{n}^{q}\in \mathbb{R}^{d_{model}\times d_{q}}$, $W_{n}^{k}\in\mathbb{R}^{d_{model}\times d_{k}}$, $W_{n}^{v}\in \mathbb{R}^{d_{model}\times d_{v}}$, $n\in(1,...,h)$, and $d_q=d_k=d_v=d_{model}/h$.
$h$ represents the number of heads for multi-head operation. 

In the encoder of the Sim-T model, the output of the Pre-MHA and Post-MHA modules will be input to the FFN module, which mainly consists of two linear layers. The first layer maps the module to the higher dimension, while the second one maps the module to the lower dimension. This operation enables the input and output of FFN to hold identical dimensions, so robust features can be efficiently extracted.

\begin{equation}
    FFN(X) = ReLU(XW_1+b_1)W_2+b_2,
\end{equation}
where $RELU$ is the activation function, $W_1\in \mathbb{R}^{d_{model}\times d_{ff}}$, $W_2\in \mathbb{R}^{d_{ff}\times d_{model}}$ are trainable parameter matrices, and $d_{ff}$ is the hidden dimension of FFN. $b_1$ and $b_2$ are bias vectors.

{\bf Decoder:} The decoder decodes the high-level acoustic features generated by the encoder. In Transformer, the output of encoder $X_{G}$ and corresponding label features $y\in \mathbb{R}^{T_{2}\times d_{model}}$ ($T_{2}$ is the length of label feature) are used as the decoder inputs. 
In the decoder, an attention mechanism is used to compute a range of context-related vector space representations of the acoustic features and label features. 
However, in the Transformer decoder, the information of the label features tends to be weakened with a deeper network layer.
Meanwhile, the decoder size significantly dominates the model decoding speed.

In the Sim-T model, we propose a new decoder structure to take advantage of the interaction relationship of label sequences, as shown on the right side of Fig. \ref{fig_model}. Each decoder layer includes four modules, that are, Pre-MHA module, MHA module, Post-MHA module and FFN module. Each module includes residual connection \cite{He2016DeepRL} and layer normalization \cite{Ba2016LayerN}. 
$Y=(Y_1,.., Y_M)$ is used to represent the output of each decoder layer, and $Y_i$ stands for the output of the $i$-th layer $(0 \leq i \leq M, Y_0 = y)$:
\begin{equation}
    Y_i = FFN(Post\mbox{-}MHA(FFN(MHA(X_G,X_G,O_i)),S_1)),
\end{equation}
\begin{equation}
    S_i, O_i = Pre\mbox{-}MHA(Y_{i-1}, Y_{i-1}, Y_{i-1}).
\end{equation}

In particular, the Post-MHA module is introduced to multiplex the attention scores in the first decoder layer, namely $S_1\in \mathbb{R}^{T_{2}\times T_{2}}$, allowing the model to learn more label interaction information. Although each decoder layer has two more modules than the original structure, the experimental results in TABLE \ref{de_tb} show that such a decoder structure only needs two layers to achieve the same effect of six layers of the Speech-Transformer model. Thus, the computational overhead and the number of parameters can be greatly reduced.

\subsection{Multiplexing Technique}\label{sec2_2}

The weight multiplexing technique proposed in MiniVit \cite{Minivit} enables all Transformer layers to share the same weights. 
Different from the scheme used in MiniVit, our proposed multiplexing technique is more flexible and can be extended to the attention score domain as it introduces a grouping operation and attention score multiplexing. This technique first divides the network layers into several groups, so layers in each group can reuse the module weights and attention scores. 
Therefore, the multiplexing technique in this study consists of two parts, that are, module weight multiplexing and attention score multiplexing.

{\bf Module Weight Multiplexing:} 
In the neural network, a model is usually composed of several stacked layers. Thus, the model size is proportional to the number of model layers. To expand the application scenarios of the model, a smaller model with similar or even better performance is desirable. To meet this demand, we explore a scheme for which the model size is determined by the setting parameters which is the number of groups in this study, rather than the number of network layers. Results show that the module multiplexing technique proposed in this study is a satisfactory solution.

In particular, module weight multiplexing has been applied in the Transformer network for model compression. As shown in Fig. \ref{fig_model}, the $L$ layers of the encoder are divided into $G$ groups. There will be $N+1$ layers in each group, including $N+1$ FFN modules, $N$ Post-MHA modules and a Pre-MHA module. Within each group, weight multiplexing is performed among the same modules. In the Pre-MHA module and Post-MHA module, the vector $V$ is generated by the linear transform $V=vW^{v}$, where $W^{v}$ is also multiplexing.
 In the decoder, the FFN module in each layer is weight multiplexing. The Post-MHA, MHA and Pre-MHA are weight multiplexing as well. The weights are only updated when being input into the next encoder group or next decoder layer.

{\bf Attention Score Multiplexing 
 :} Global context capture of the sequences using the attention mechanism is an important means for the Transformer model to achieve excellent performance. As mentioned in Section \ref{sec2_1}, the attention score multiplexing can save calculations and impose little impact on model performance. Therefore, a scheme called attention score multiplexing has been proposed and applied in this study.

In the encoder, the attention scores are computed only once in each group in the Pre-MHA module. After that they are shared among the Post-MHA modules in the following $N$ layers. In this process, the $Q=qW^{q}$, $K=kW^{k}$ and $S$ computations can be reduced $N$ times. Therefore, the Post-MHA in the $j$-th layer of the $i$-th encoder group can be expressed as:

\begin{equation}
    Post\mbox{-}MHA(S_i,X_{i}^{j-1}) = S_iV = S_iX_{i}^{j-1}W^{v}.
\end{equation}

The application of attention score multiplexing in the decoder is different from that used in the encoder. The main purpose is to reuse the interaction relationship $S_{1}(Q,K)= SOFTMAX[({{y}{y^{T}}})/{\sqrt{d_k}}]$ of the label features $y$. The decoder includes $M$ Post-MHA modules, all of which share the same attention score $S_{1}$. Another input $v_i=MHA(X_G,O_i,X_G)$ is the high-level acoustic features. Therefore, the Post-MHA module in the $i$-th decoder layer can be expressed as:
\begin{equation}
    Post\mbox{-}MHA(S_1,v_i) = S_1V = S_1v_{i}W^{v}.
\end{equation}

Under the help of the attention score multiplexing technique, each decoder layer will contain the interactive information of label features. This guarantees that robust features, which are useful for model improvement, can be effectively learned.

\section{Experiment}




\subsection{Experimental Setting}\label{sec4_1}

Three widely-used datasets, that are Aishell-1 \cite{Aishell1}, HKUST \cite{liu2006hkust} and WSJ \cite{paul-baker-1992-design}, are selected to validate the effectiveness of the proposed Sim-T network. It should be noted that the Aishell-1 and HKUST are Chinese datasets and WSJ is an English dataset. Fig. \ref{fig_model} shows the effective preprocessing for the input audio signal at the front-end, which includes framing, windowing, fast fourier transform and discrete cosine transform. The window size is 25ms. The window shift is 10ms, and the output feature is an 80-dimensional log Mel-filterbank.

All experiments are conducted by the employment of the ESPnet toolkit\cite{watanabe2018espnet}. Similar as other approaches implemented in ESPnet, the Sim-T model has been trained for 50 epochs on the Aishell-1 dataset, 35 epochs on the HKUST dataset and 100 epochs on the WSJ dataset. Label smoothing and dropout regulation ($p=0.1$) are adopted to prevent over-fitting. The optimizer is Adam \cite{2014Adam} with a learning rate of $0.002$, $25000$ warm-up steps, $\epsilon$ at $10^{-9}$, $\beta1=0.9$ and $\beta2=0.98$. The other parameters are: $d_{model}=256$, $d_{ff}=2048$, $h=4$, $N=12$ and $M=2$. The parameters associated with attention dimensions are set as: $d_ {q} = d_ {k} = d_ {v} = d_ {model}/h = 64$. The initial value of each batch is set as 64, and the $beam\_size$ in the beam search algorithm is 10. The language model (LM) contains 16 layers and is trained for 15 epochs with the Transformer framework.

\begin{table}[tp]
	\centering
		\caption{Comparison among Sim-T and other ASR models in Aishell-1 dataset} 
		\vspace{2pt}
		\setlength{\tabcolsep}{3mm}{
	\begin{tabular}{lcc}
		\hline
    		Model&Dev (\%) &Test (\%) \\
		
		\hline
		Transformer without LM \cite{speech_transformer}       & 5.5        & 5.9       \\
		Transformer with LM \cite{speech_transformer}       & 5.3        & 5.6       \\
		\hline
        RNN-T \cite{Transducers}  &  10.13      &  11.82     \\ 
        SA-T \cite{Transducers}   &  8.30       &  9.30      \\
        Chunk-Flow SA-T \cite{Transducers}  & 8.58        & 9.80       \\
        Sync-Transformer \cite{Tian2020SynchronousTF}   & 7.91        &  8.91     \\
		Masked-NAT \cite{chen2020non}  &      6.4&   7.1\\
		ESPNet-RNN \cite{20dsad19} & 6.8 & 8.0 \\
		Insertion-NAT \cite{2353803bf0d24fddaf55abd105215289} & 6.1  &  6.7\\
		LASO \cite{bai2020listen}          &5.8 &  6.4\\
        AT \cite{9413429}&   5.5  &   5.9 \\
        Realformer with LM \cite{he2020realformer} & 5.7 &6.1 \\
    	\hline
    	Sim-T without LM                     & 5.3 & 5.7     \\
    	Sim-T with LM                               & \textbf{4.9} & \textbf{5.2}     \\
		\hline\\
	\end{tabular}}
	\label{aishell_tb}
	\vspace{-15pt}
\end{table} 

\begin{table}[tp]
	\centering
		\caption{Performance of ASR models in HKUST dataset.} 
		\vspace{-2pt}
	\begin{tabular}{lcc}
		\hline
		Model & CER (\%) \\
		
		\hline
        Chain-TDNN \cite{povey16_interspeech} & 23.7 \\
        Self-attention Aligner \cite{Dong2019SelfattentionAA} & 24.1 \\
        Extended-RNA \cite{dong2019extending} & 26.6 \\
        Joint CTC-attention model/ESPNet \cite{7953075} & 27.4 \\
		SAM with LM \cite{Dong2019SelfattentionAA}& 24.92\\
        CTC with LM \cite{7472152} & 34.8 \\
        Transformer without LM \cite{speech_transformer} & 21.7\\
        Transformer with LM \cite{speech_transformer} & 21.5\\
        \hline
        Sim-T without LM & 21.6 \\
        Sim-T with LM & \textbf{21.2} \\
        \hline\\
	\end{tabular}
	\label{hkust_tb}
	\vspace{-10pt}
\end{table} 

\begin{table}[tp]
	\centering
		\caption{WER comparison of ASR models in WSJ dataset.} 
		\vspace{-2pt}
	\begin{tabular}{lcc}
		\hline
		Model & dev93 (\%) & eval92 (\%) \\
		
		\hline
           DeCoAR \cite{9053176} & 8.3 &  4.6  \\
           SAN-CTC\cite{SAN-CTC} & 8.9 &  5.9  \\
           E2E-SincNet \cite{9053954}& 7.8 &  4.7 \\
           Transformer with LM \cite{speech_transformer} & 6.6 & 4.6   \\
        \hline
        Sim-T with LM & \textbf{6.5} & \textbf{4.4} \\
        \hline\\
	\end{tabular}
	\label{wsjtb}
	\vspace{-10pt}
\end{table} 
\subsection{Comparison with Other Popular Models}\label{sec4_2}

As shown in TABLE \ref{aishell_tb}, it can be clearly seen that the proposed Sim-T model gives better performance than other popular approaches on the Aishell-1 dataset. When LM is adopted, the Sim-T achieves 4.9\% and 5.2\% CER on the \textit{Dev} and \textit{Test} datasets, respectively. Both of them have 0.4\% CER improvement in comparison with the baseline Transformer model. When LM is excluded, the achieved CER on \textit{Dev} and \textit{Test} are 5.3\% CER and 5.7\% CER, respectively, and they are still 0.2\% better than the baseline. Notably, the number of parameters for Sim-T is only 15.62M, so a reduction of 14.74M (48\%) parameters can be obtained compared with 30.36M in the baseline Transformer model.

The experimental results on the HKUST dataset are shown in TABLE \ref{hkust_tb}. Some popular results are also presented for comparison. Apparently, one can find that the Sim-T model gives better performance and less parameters and calculations are involved. When LM is included and excluded, the CER of Sim-T are 21.2\% and 21.6\%, respectively, which are 0.3\% and 0.1\% better than the baseline Transformer network. Here, the size of the baseline Transformer model is 29.85M and for Sim-T model is 17.77M, so the parameter number shows up a reduction of 40\%. Also, Sim-T gives better performance than other popular networks.

In addition, further experiments on the English dataset named WSJ have been conducted to verify the performance of the Sim-T network in different languages, and results are shown in TABLE \ref{wsjtb}. Compared with popular networks, Sim-T can also achieve certain improvements with fewer parameters. The word error rate (WER) on \textit{eval93} and \textit{eval92} are 6.5\% and 4.4\%, respectively, with improvements of 0.1\% and 0.2\% when in comparison with the baseline Transformer. Here, the model size for the baseline Transformer is 27.14M, but it drops to 16.28M for the Sim-T with 40\% parameters reduction.

On the basis of the above experimental results, conclusions can be drawn that the multiplexing technique is useful for network compression. Moreover, the network performance is surprisingly developed instead of degraded with appropriate parameter settings. In particular, when the multiplexing technique is adopted into the Transformer network, the parameter amount shows up at least 40\% reduction without model degradation.

\begin{table}[tp]
	\centering
		\caption{Comparison of Sim-T\_$6\times4$, Sim-T\_$6\times2$ and baseline Transformer.} 
	\begin{tabular}{lcccc}
		\hline
		Model & Dev(\%) & Test(\%) & Model Size($M$) \\
		
		\hline
		Transformer with LM & 5.3 & 5.6 & 30.36 \\
        Transformer without LM & 5.5 & 5.9 & - \\
        \hline
		Sim-T\_$6\times4$ with LM & 4.9 & 5.3 & 18.25 ($41\%\downarrow$) \\
		Sim-T$\_6\times4$  without LM & 5.5 & 5.9 & - \\
       \hline
		Sim-T$\_6\times2$  with LM & \textbf{4.9} & \textbf{5.2} & \textbf{15.62} ($48\%\downarrow$) \\
		Sim-T$\_6\times2$  without LM & 5.3 & 5.7 & - \\
		\hline\\
	\end{tabular}
	\label{de_tb}
	\vspace{-10pt}
\end{table}

\subsection{Ablation Studies}\label{sec4_4}

In order to systematically examine the effectiveness of the proposed Sim-T model, investigations on major components of the network have been carried out and all the experiments were conducted on the Aishell-1 dataset. 
Some items are worth being explored for model understanding, such as the superiority of the Sim-T network structure, the effect of multiplexing and grouping in the proposed Sim-T network, and the extended application of multiplexing. 

{\bf Superiority of Sim-T Model Structure: } 
Three models are selected for experimental exploration to prove the high performance of Sim-T model in the case of compressed models when using the same or less number of modules as the Transformer model does.
These chosen models are the baseline Transformer model with twelve encoder layers and six decoder layers, the Sim-T\_$6\times4$ model of six encoder groups and four decoder layers, and Sim-T$\_6\times2$ model of six encoder groups and two decoder layers, respectively. There are two encoder layers in each group of Sim-T\_$6\times4$ and Sim-T$\_6\times2$. 
It is worth mentioning that in Transformer, each decoder layer has one FFN module and two MHA-related modules, while in Sim-T, each decoder layer has two FFN modules and three MHA-related modules.
To keep the fairness, Sim-T$\_6\times4$ is selected for comparison with the Transformer network since they hold similar number of corresponding modules. 
The mentioned decoder in Sim-T$\_6\times2$  has fewer modules, which is used to verify the superiority of this new decoder structure proposed in this paper.

The corresponding experimental results are given in TABLE \ref{de_tb}. Based on the experimental results, one can note that when the LM is employed, the Sim-T\_$6\times4$ improves the model performance by 0.4\% and 0.3\% CER on the \textit{Dev} and \textit{Test} datasets, respectively, compared with the baseline Transformer. Also, the model gives the same performance as the baseline when LM is excluded. It is of great importance to point out that the model size of Sim-T\_$6\times4$ is 18.25M and the baseline Transformer is 30.36M, demonstrating 41\% parameters in the baseline have been compressed. 
Based on the above analysis, one can notice that the proposed Sim-T network structure is superior to the Transformer since better performance can be obtained with similar number of corresponding modules.

Noticeably, the Sim-T$\_6\times2$ gives better performance and is more lightweight in comparison with the Sim-T\_$6\times4$. In contrast, Sim-T$\_6\times2$ reduces the amount of model parameters by 0.7\% compared to Sim-T$\_6\times4$. When using LM, Sim-T$\_6\times2$ on the \textit{Dev} dataset gives 0.1\% CER improvement over Sim-T$\_6\times4$. When LM is excluded, Sim-T$\_6\times2$, on \textit{Dev} and \textit{Test} datasets, improves CER by 0.2\% compared to Sim-T$\_6\times4$. In addition, Sim-T$\_6\times2$ improves the model performance by 0.4\% CER on both \textit{Dev} and \textit{Test} compared with the baseline Transformer with LM. When LM is not applied, 0.2\% CER improvement can be achieved on both \textit{Dev} and \textit{Test}. The model size of Sim-T$\_6\times2$ is 15.62M, where 48\% parameters can be compressed when it is compared with the baseline. 
Surprisingly, the performance of Sim-T$\_6\times2$ is better than the Sim-T$\_6\times4$, and this further proves that the proposed network structure is useful for module compression.
\begin{table}[tp]
	\centering
		\caption{The effect on weight multiplexing of LN on model performance and model size.} 
	\begin{tabular}{lcccc}
		\hline
		Model & Dev(\%) & Test(\%) & Model Size($M$) \\
		
		\hline
		Transformer with LM & 5.3 & 5.6 & 30.36 \\
        \hline
		Sim-T\_\Rmnum{1} with LM & 4.9 & 5.2 & 15.62 ($48\%\downarrow$) \\
		Sim-T\_\Rmnum{2} with LM & 5.2 & 5.6 & \textbf{9.32} 
        ($\textbf{69\%}\downarrow$) \\
        \hline
	\end{tabular}
	\label{LN12_tb}
	\vspace{-10pt}
\end{table} 

\begin{table}[!htbp]
    \centering
    \caption{The multiplexing technique is used in different part of Transformer network.}
    \begin{tabular}{ccccc}
    \hline
     Encoder&Decoder  &Dev(\%) &Test(\%) & Model Size(M)  \\
    \hline
    \texttimes &\texttimes  &5.3  &5.6 & 30.36\\
    \checkmark & \texttimes & 5.3 &5.5 & 17.21 ($43\%\downarrow$)\\
   \texttimes &\checkmark  &5.2  &5.6 & 22.47 ($26\%\downarrow$)\\
   \checkmark  &\checkmark  &5.2  &5.6 & 9.32 ($69\%\downarrow$)\\
   \hline
    \end{tabular}
    \label{de1_tb}
    \vspace{-5pt}
\end{table}
{\bf Multiplexing Technique: } As mentioned in section \ref{sec2_1}, each module in the encoder and decoder of Sim-T includes residual connection and layer normalization. Here we conduct an in-depth discussion on how the model performance and size are affected by parameter sharing of LNs in the same type of modules. 
For example, in each encoder group, the LN parameters are shared between the Pre-MHA module and Post-MHA module. They are also shared in each FFN module. A similar procedure is conducted in the decoder to further reduce the parameter amount. 
The networks in which the LN without and with the implementation of the multiplexing technique are called Sim-T\_\Rmnum{1} and Sim-T\_\Rmnum{2}, respectively.
The difference between the above two versions has been verified by experiments, and results are shown in TABLE \ref{LN12_tb}. One can see that although Sim-T\_\Rmnum{2} has a 0.1\% CER improvement on both \textit{Dev} dataset when LM is involved, the model size of Sim-T\_\Rmnum{2} is only 9.32M, which is 69\% lower than the baseline Transformer. 
That is to say, more parameters can be compressed on the basis of Sim-T\_\Rmnum{1}. The experiments indicate that extension of the multiplexing technique is likely to make the model to be more lightweight while its performance can be maintained or even improved.

Moreover, the proposed multiplexing technique has also been applied in baseline Transformer model to verify its impact on the encoder and decoder, respectively. The experimental results are shown in TABLE \ref{de1_tb}. When only applying the multiplexing technique in the encoder, the CER can be improved by 0.1\% over the baseline Transformer on the \textit{Test} dataset with a model size of 17.21M and a 43\% compression. On the other hand, 0.1\% CER improvement over the baseline Transformer on the \textit{Dev} dataset with a model size of 22.47M and a 26\% compression can be obtained if only the decoder has been applied the multiplexing technique. Implementation of the weight reuse technique in both the encoder and decoder can compress the model size to 9.32M and improve 0.1\% CER on the \textit{Dev} dataset. Therefore, it is logic to say that the model can be compressed with no sacrifice of the model performance regardless of the implementation of the multiplexing technique in either encoder or decoder.

\begin{table}[tp]
	\centering
		\caption{Comparison of different sizes of encoder groups.} 
	\begin{tabular}{lcccc}
		\hline
		Model & Dev(\%) & Test(\%) & Model Size($M$) \\
		
		\hline
		Transformer with LM & 5.3 & 5.6 & 30.36 \\
        \hline
		Sim-T$\_6\times2$  with LM & 5.2 & 5.6 & 9.32 ($69\%\downarrow$) \\
		Sim-T$\_4\times2$  with LM & 5.3 & 5.6 & 8.97 ($71\%\downarrow$) \\
  	Sim-T$\_3\times2$  with LM & 5.5 & 5.8 & 8.52 ($71\%\downarrow$) \\
		Sim-T$\_2\times2$  with LM & 5.7 & 6.1 & 8.28 ($72\%\downarrow$) \\
		Sim-T$\_1\times2$  with LM & 6.1 & 6.6 & 7.99 ($73\%\downarrow$) \\
		\hline\\
	\end{tabular}
	\label{engroup_tb}
	\vspace{-10pt}
\end{table} 

{\bf Analysis of Group Quantities: }The cross-layer weight sharing technique used in ALBert \cite{ALBert} simply shares the weights of all layers in the model. A similar technique is also adopted in Universal Transformer \cite{dehghani2018universal} and MiniVit \cite{Minivit}. Different from these two approaches, the Sim-T proposed in this study divides all encoder layers into multiple groups, and the proposed multiplexing technique is used in each group. For example, twelve encoder layers are divided into four groups, so corresponding modules in each group multiplexes the module weights and attention scores. 
When there is only one encoder group, it means that weights from all layers are shared.
Given that the Sim-T$\_i\times2$ means that all encoder layers are divided into $i$ groups and two decoder layers are included.
Here, we choose Sim-T\_\Rmnum{2} for experimental verification, and results are shown in TABLE \ref{engroup_tb}. Based on the experimental results, the encoder group is gradually reduced from six to one, while the model performance on \textit{Dev} and \textit{Test} gradually decreases from 5.2\% to 6.1\% CER and from 5.6\% to 6.6\% CER, respectively. 
It should be highlighted that the model performance degrades with the decreasing number of encoder groups. 
Also, the model size gradually decreases from 9.32M to 7.99M with the increasing number of encoder groups. From the table, one can note that the model performance is undesirable when the encoder group is one, although a greater degree of model compression can be obtained. Therefore, setting an appropriate number of groups is of an utmost important trade-off between the model performance and size when the multiplexing technique is applied for compression purpose. For example, the most applicable encoder group number in this study is six. 

\begin{table}[tp]
	\centering
		\caption{The model size and memory of LM and SimLM.} 
	\begin{tabular}{lcccc}
		\hline
		Model  & PPL & Model Size($M$) & Memory($G$) \\
		
		\hline
		LM \cite{transformer_LM} & 49.74 & 53.24 & 1.396 \\
        SimLM & 51.86 & \textbf{5.94}($\textbf{89\%}\downarrow$) & \textbf{0.572}($\textbf{59\%}\downarrow$)\\
		\hline\\
	\end{tabular}
	\label{lm1_tb}
	\vspace{-10pt}
\end{table}

\begin{table}[tp]
	\centering
		\caption{Performance comparison of Transformer with LM and with SimLM on the Aishell-1 dataset.} 
	\begin{tabular}{lcccc}
		\hline
		Model & Dev(\%) & Test(\%) \\
		
		\hline
		Transformer with LM & 5.3 & 5.6  \\
        Transformer with SimLM & 5.8 & 6.1 \\
		\hline\\
	\end{tabular}
	\label{lm2_tb}
	\vspace{-10pt}
\end{table} 

{\bf Application of Language Model: }Finally, extended experiments on the language model have been conducted 
to further validate the compression effect of the proposed multiplexing technique. The baseline used for comparison is a Transformer-based language model (LM) consisting of sixteen encoder layers. This study proposes a new language model called SimLM in which the multiplexing technique is applied for the simplification of LM. The SimLM divides sixteen encoder layers into eight encoder groups, and each of them contains two layers. The perplexity (PPL), model size and memory size during training of the two language models are shown in TABLE \ref{lm1_tb}. Among them, the PPL of LM is 49.74, and it increases to 51.86 (with 2.12 increment) for the PPL of SimLM. However, the model size of LM is 53.24M, while the value is only 5.94M for SimLM. 
That is to say, 89\% parameters have been compressed. The memory size of LM is 1.396G, but it drops to 0.572G for SimLM, indicating  59\% of memory space has been saved. 
In addition, both LM and SimLM are inserted into the Speech Transformer network for ASR task testing to show the contribution of the proposed SimLM. The experimental results are shown in TABLE \ref{lm2_tb}.
On \textit{Dev} and \textit{Test}, the performance of the Transformer model with SimLM is 0.5\% CER lower than that with LM. 
From the table, one can note that there is an attractive trade-off between the model performance and parameter number. 
The reason is that model size and memory size footprint are significantly decreased while the degradation of model performance is always negligible. This further proves that the proposed multiplexing technique can compress the model to a large extent without the serious sacrifice of the model performance.

On the basis of all the ablation studies mentioned above, the proposed Sim-T has been explored and validated in multiple technical issues. The conclusions can be drawn as follows: First, applying interaction among label features to each decoder layer is useful for model improvement. In this way, better performance with fewer decoder layers than the baseline Transformer network can be obtained. Second, parameter multiplexing in LN among modules can further compress the model. Third, the model size becomes smaller if the number of encoder groups is decreasing, but model performance tends to degrade. 
Last, the multiplexing technique can be extended to language models, which can greatly compress the model without causing serious effects on the model performance.

\section{Conclusion}
A new lightweight ASR model called Sim-T, which can efficiently simplify the Transformer network using the multiplexing technique, has been proposed in this study. A combination of module weight multiplexing, attention score multiplexing, and a new decoder structure allows Sim-T to efficiently reduce parameters and give better performance than the original Transformer model. Its usefulness has been verified through extensive experiments on three widely used datasets, that are, Aishell-1, HKUST, and WSJ. In addition, the weight multiplexing technique has also been applied to the language model to verify its generalization ability. Results show that Sim-T is able to reduce at least 40\% of the number of model parameters with almost no performance degradation.

\section*{Acknowledgments}
This work was supported by Research Foundation No.2020A1515111107, No.2021B1515120025 and No.2021KSYS008.

\vfill


\begin{thebibliography}{1}
\bibliographystyle{IEEEtran}

\bibitem{transformer2017}Vaswani A, Shazeer N, Parmar N, et al. ``Attention is all you need'', {\em Advances in Neural Information Processing Systems}, 2017, pp. 5999-6009.

\bibitem{Bert}Devlin J, Chang M W, Lee K, et al. ``BERT: Pre-training of Deep Bidirectional Transformers for Language Understanding'', {\em  North {A}merican Chapter of the Association for Computational Linguistics}, 2019, pp. 4171-4186.


\bibitem{ALBert}Lan Z, Chen M, Goodman S, et al. ``ALBERT: A Lite BERT for Self-supervised Learning of Language Representations'', {\em International Conference on Learning Representations}, 2020.


\bibitem{Q8Bert}Zafrir O, Boudoukh G, Izsak P, et al. ``Q8bert: Quantized 8bit bert'', {\em Fifth Workshop on Energy Efficient Machine Learning and Cognitive Computing-NeurIPS Edition (EMC2-NIPS)}, 2019, pp. 36-39.

\bibitem{EAPT}Lin X, Sun S, Huang W, et al. ``EAPT: efficient attention pyramid transformer for image processing'', {\em IEEE Transactions on Multimedia}, vol.25, 2021, pp. 50-61.



\bibitem{liu2021swin}Liu Z, Lin Y, Cao Y, et al. ``Swin transformer: Hierarchical vision transformer using shifted windows'', {\em Proceedings of the IEEE/CVF International Conference on Computer Vision}, 2021, pp. 10012-10022.


\bibitem{fan2021multiscale}Fan H, Xiong B, Mangalam K, et al. ``Multiscale vision transformers'', {\em Proceedings of the IEEE/CVF International Conference on Computer Vision}, 2021, pp. 6824-6835.

\bibitem{transformer_XL}Dai Z, Yang Z, Yang Y, et al. ``Transformer-XL: Attentive Language Models beyond a Fixed-Length Context'', {\em Proceedings of the 57th Annual Meeting of the Association for Computational Linguistics}, 2019, pp. 2978-2988.

\bibitem{transformer_LM}Al-Rfou R, Choe D, Constant N, et al. `` Character-level language modeling with deeper self-attention'', {\em AAAI conference on artificial intelligence}, 2019, pp. 3159-3166.

\bibitem{speech_transformer}Dong L, Xu S, Xu B. ``Speech-transformer: a no-recurrence sequence-to-sequence model for speech recognition'', {\em IEEE international conference on acoustics, speech and signal processing (ICASSP)}, 2018, pp. 5884-5888.

\bibitem{LFEformer}Wei G, Duan Z, Li S, et al. ``LFEformer: Local Feature Enhancement Using Sliding Window with Deformability for Automatic Speech Recognition'', {\em IEEE Signal Processing Letters}, vol. 30, 2023, pp. 180-184.


\bibitem{gulati2020conformer}Gulati A, Qin J, Chiu C C, et al. ``Conformer: Convolution-augmented transformer for speech recognition'', {\em Interspeech}, 2020, pp. 5036-5040.

\bibitem{he2020realformer}He R, Ravula A, Kanagal B, et al. ``RealFormer: Transformer Likes Residual Attention'', {\em ACL-IJCNLP}, 2021, pp. 929-943.

\bibitem{lstransformer}Li J, Wang X, Li Y. ``speechtransformer for large-scale mandarin chinese speech recognition'', in {\em ICASSP}, 2019, pp. 7095-7099.

\bibitem{aes} Zhikui Duan, Guozhi Gao, Jiawei Chen, et al. ``Dual-Residual Transformer Network for Speech Recognition'', {\em journal of the audio engineering society}, 70(10), 2022, pp. 871-881.

\bibitem{lecun1989generalization}LeCun Y. ``Generalization and network design strategies'', {\em neural computation}, 1989, pp. 143-155.

\bibitem{nowlan2018simplifying}Nowlan S J, Hinton G E. ``Simplifying neural networks by soft weight sharing'', {\em Neural Computation}, 1992, pp. 473-493.

\bibitem{dehghani2018universal}Dehghani M, Gouws S, Vinyals O, et al. ``Universal transformers'', {\em ICLR}, 2019.

\bibitem{Redundancy}Dalvi F, Sajjad H, Durrani N, et al. ``Analyzing Redundancy in Pretrained Transformer Models'', {\em  Empirical Methods in Natural Language Processing (EMNLP)}, 2020, pp. 4908-4926.

\bibitem{shim2021understanding}Shim K, Choi J, Sung W. ``Understanding the role of self attention for efficient speech recognition'', {\em International Conference on Learning Representations}, 2021.


\bibitem{tmm_speech}Tao F, Busso C. ``End-to-end audiovisual speech recognition system with multitask learning'', {\em IEEE Transactions on Multimedia}, vol. 23, 2020, pp. 1-11.




\bibitem{CTC}Graves A, Fernández S, Gomez F, et al. ``Connectionist temporal classification: labelling unsegmented sequence data with recurrent neural networks'', {\em Proceedings of the 23rd international conference on Machine learning}, 2006, pp. 369-376.


\bibitem{RNN}Graves A, Mohamed A, Hinton G. ``Speech Recognition with Deep Recurrent Neural Networks'', {\em IEEE International Conference on Acoustics, Speech and Signal Processing}, 2013, pp. 6645-6649.


\bibitem{LSTM}Malhotra P, Vig L, Shroff G, et al. ``Long short term memory networks for anomaly detection in time series'', {\em ESANN}, 2015, pp. 89-94.

\bibitem{LAS}Chan W, Jaitly N, Le Q V, et al. ``Listen, attend and spell: A neural network for large vocabulary conversational speech recognition'', {\em IEEE international conference on acoustics, speech and signal processing (ICASSP)}, 2016, pp. 4960-4964.

\bibitem{kovaleva2019revealing}Kovaleva O, Romanov A, Rogers A, et al. ``Revealing the dark secrets of BERT'', {\em Empirical Methods in Natural Language Processing and the 9th International Joint Conference on Natural Language Processing (EMNLP-IJCNLP)}, 2019, pp. 4365-4374.

\bibitem{Tinybert}Jiao X, Yin Y, Shang L, et al. ``TinyBERT: Distilling {BERT} for Natural Language Understanding'', in {\em Findings of the Association for Computational Linguistics: EMNLP 2020}, 2020, pp. 4163–4174.

\bibitem{DeiT}Touvron H, Cord M, Douze M, et al. ``Training data-efficient image transformers \& distillation through attention'', {\em International conference on machine learning}, 2021, pp. 10347-10357.


\bibitem{Dynamicvit}Rao Y, Zhao W, Liu B, et al. ``Dynamicvit: Efficient vision transformers with dynamic token sparsification'', {\em Advances in neural information processing systems}, 2021, pp. 13937-13949.









\bibitem{Leve_transformer}Gu J, Wang C, Zhao J. ``Levenshtein transformer'', {\em Advances in Neural Information Processing Systems}, 2019.

\bibitem{Minivit}Zhang J, Peng H, Wu K, et al. ``Minivit: Compressing vision transformers with weight multiplexing'', {\em Conference on Computer Vision and Pattern Recognition}, 2022, pp. 12145-12154.

\bibitem{He2016DeepRL}He K, Zhang X, Ren S, et al. ``Deep Residual Learning for Image Recognition'', in {\em IEEE Conference on Computer Vision and Pattern Recognition}, 2016, pp. 770-778.

\bibitem{Ba2016LayerN}Ba J L, Kiros J R, Hinton G. ``Layer Normalization'', {\em ArXiv:1607.06450}, 2016.


\bibitem{Aishell1}Bu H, Du J, Na X, et al. ``Aishell-1: An open-source mandarin speech corpus and a speech recognition baseline'', {\em Conference of the Oriental Chapter of the International Coordinating Committee on Speech Databases and Speech I/O Systems and Assessment}, 2017, pp. 1-5.

\bibitem{liu2006hkust}Liu Y, Fung P, Yang Y, et al. ``Hkust/mts: A very large scale mandarin telephone speech corpus'', {\em International Symposium on Chinese Spoken Language Processing}, 2006, pp. 724-735.

\bibitem{paul-baker-1992-design}Paul D B, Baker J. ``The design for the Wall Street Journal-based CSR corpus'', {\em Proceedings of the workshop on Speech and Natural Language}, 1992, pp. 357–362.

\bibitem{watanabe2018espnet}Watanabe S, Hori T, Karita S, et al. ``{ESPnet}: End-to-End Speech Processing Toolkit'', {\em Interspeech}, 2018, pp. 2207-2211.

\bibitem{2014Adam}Kingma D, Ba J L. ``Adam: A Method for Stochastic Optimization'', {\em International Conference for Learning Representations}, 2015.


\bibitem{Transducers}Tian Z, Yi J, Tao J, et al. ``Self-Attention Transducers for End-to-End Speech Recognition'', {\em Interspeech}, 2019, pp. 4395-4399.


\bibitem{Tian2020SynchronousTF}Tian Z, Yi J, Bai Y, et al. ``Synchronous Transformers for end-to-end Speech Recognition'', {\em IEEE International Conference on Acoustics, Speech and Signal Processing}, 2020, pp. 7884-7888.

\bibitem{chen2020non}Chen N, Watanabe S, Villalba J, et al. ``Non-autoregressive transformer for speech recognition'', {\em IEEE Signal Processing Letters}, 2021, vol. 28, pp. 121-125.

\bibitem{20dsad19}Karita S, Chen N, Hayashi T, et al. ``A Comparative Study on Transformer vs RNN in Speech Applications'', {\em IEEE Automatic Speech Recognition and Understanding Workshop}, 2019, pp. 449-456.

\bibitem{2353803bf0d24fddaf55abd105215289}Fujita Y, Watanabe S, Omachi M, et al. ``Insertion-based modeling for end-to-end automatic speech recognition'', {\em Interspeech}, 2020, pp. 3660-3664.

\bibitem{bai2020listen}Bai Y, Yi J, Tao J, et al. ``Listen Attentively, and Spell Once: Whole Sentence Generation via a Non-Autoregressive Architecture for Low-Latency Speech Recognition'', {\em Interspeech}, 2020, pp. 3381-3385.

\bibitem{9413429}Fan R, Chu W, Chang P, et al. ``CASS-NAT: CTC Alignment-Based Single Step Non-Autoregressive Transformer for Speech Recognition'', {\em IEEE International Conference on Acoustics, Speech and Signal Processing}, 2021, pp. 5889-5893.

  
\bibitem{povey16_interspeech}Povey D, Peddinti V, Galvez D, et al. ``Purely Sequence-Trained Neural Networks for ASR Based on Lattice-Free MMI'', {\em Interspeech}, 2016, pp. 2751-2755.



\bibitem{Dong2019SelfattentionAA}Dong L, Wang F, Xu B, ``Self-attention Aligner: A Latency-control End-to-end Model for ASR Using Self-attention Network and Chunk-hopping'', {\em IEEE International Conference on Acoustics, Speech and Signal Processing}, 2019, pp. 5656-5660.

\bibitem{dong2019extending}Dong L, Zhou S, Chen W, et al. ``Extending Recurrent Neural Aligner for Streaming End-to-End Speech Recognition in Mandarin'', {\em Interspeech}, 2018, pp. 816-820.

\bibitem{7953075}Kim S, Hori T, Watanabe S, et al. ``Joint CTC-attention based end-to-end speech recognition using multi-task learning'',  {\em IEEE International Conference on Acoustics, Speech and Signal Processing}, 2017, pp. 4835-4839.


  
\bibitem{7472152}Miao Y, Gowayyed M, Na X, et al. ``An empirical exploration of CTC acoustic models'', {\em IEEE International Conference on Acoustics, Speech and Signal Processing}, 2016, pp. 2623-2627.

\bibitem{9053176}Ling S, Liu Y, Salazar J, et al. ``Deep Contextualized Acoustic Representations for Semi-Supervised Speech Recognition'', {\em IEEE International Conference on Acoustics, Speech and Signal Processing}, 2020, pp. 6429-6433.



\bibitem{SAN-CTC}Salazar J, Kirchhoff K, Huang Z. ``Self-attention networks for connectionist temporal classification in speech recognition'', {\em IEEE International Conference on Acoustics, Speech and Signal Processing}, 2019, pp. 7115-7119.


\bibitem{9053954}Parcollet T, Morchid M, Linares G. ``E2E-SINCNET: Toward Fully End-To-End Speech Recognition'', {\em IEEE International Conference on Acoustics, Speech and Signal Processing}, 2020, pp. 7714-7718.




\end{thebibliography}
\end{document}